\documentclass[prb,
                    preprint
                    ]{revtex4}

\usepackage{epsfig}
\usepackage{amsmath}
\usepackage{subfigure}

\begin{document}

\def\e{\begin{equation}}
\def\f{\end{equation}}
\def\l#1{\label{eq:#1}}
\def\r#1{(\ref{eq:#1})}

\title{A three-dimensional backward-wave network matched with free space}

\author{Pekka~Alitalo, Olli~Luukkonen and Sergei~Tretyakov}

\affiliation{TKK Helsinki University of Technology, Radio Laboratory / SMARAD Centre of Excellence\\
P.O. Box 3000, FI-02015 TKK, Finland\\
{\rm E-mails: pekka.alitalo@tkk.fi, olli.luukkonen@tkk.fi,
sergei.tretyakov@tkk.fi}}

\date{\today}

\begin{abstract}

A backward-wave slab based on a capacitively and inductively
loaded three-dimensional transmission-line network is designed in
such a way that impedance-matching with free space is obtained. To
enable field propagation from free space to the network and vice
versa, the use of a transition layer is proposed. Matching of the
designed network with free space and negative refraction occurring
at the slab interfaces are confirmed with full-wave simulations.

\end{abstract}

\maketitle

\section{Introduction}

As suggested by Veselago,\cite{Veselago} a material with negative
permittivity $\varepsilon$ and permeability $\mu$ (a backward-wave
material) can be used as a flat lens that focuses propagating
electromagnetic waves. Due to the negative $\varepsilon$ and
$\mu$, the wave propagation in this material differs significantly
from any material found in nature, since the phase and group
velocities are antiparallel. Furthermore, as shown by
Pendry,\cite{Pendry} a slab of such material can be used as a
superlens that, besides focusing the propagating waves, enhances
the evanescent waves of a source.

First experimental demonstration of negative refraction, that
occurs on the interface between a material with positive
$\varepsilon$ and $\mu$ and a material with negative $\varepsilon$
and $\mu$, was achieved with a slab made of a composite material
consisting of metal wires and split-ring-resonators.\cite{Shelby}
The use of resonant particles in creating the wanted negative
$\mu$ has the drawback of very narrow operation bandwidth and high
sensitivity to losses. An alternative approach to creating a
``material'' with negative $\varepsilon$ and $\mu$, based on
loaded transmission lines (TLs), has been
proposed.\cite{Eleftheriades,Caloz} The benefit of this approach
is the fact that the exotic wave propagation is not due to use of
resonant particles and thus the operational bandwidth and losses
are not so critical issues. The drawback of such structures in
superlens applications is that coupling of waves from free space
to such a network is not trivial. Indeed, superlenses proposed in
the literature that are based on the TL-method, have used sources
which are embedded in a TL network as well, see e.g. Ref.~6.
Recently, also three-dimensional extensions of the TL-method have
been proposed\cite{Grbic2,Hoefer,Alitalo1} and
realized.\cite{Alitalo2}

Recently, a design of a TL network with negative index of
refraction, that can be matched with free space, was
proposed.\cite{Iyer} This approach can be realized for
two-dimensional TL networks, i.e., a set of two-dimensional TL
networks can be stacked on top of each other creating a volumetric
slab. In this letter, we propose a transition layer to couple
waves from free space to a TL network such as proposed in Ref.~6
(two-dimensional TL network) and in Ref. 9 (three-dimensional TL
network). The transition layer is effectively an array of antennas
that covers the whole interface between free space and the TL
network. This approach, as compared to the previous
design,\cite{Iyer} has the benefit of freedom in the design of the
TL network, since the network itself does not have to be coupled
with free space. Moreover, there are no parasitic forward waves,
since fields are concentrated in the TLs only.

\section{Design of the loaded transmission-line network}

In this letter we study the structure presented in Refs.~9 and 10,
although the proposed method of matching a TL network with free
space can be used for other types of networks as well.\cite{cloak}
As compared to the previous designs,\cite{Alitalo1,Alitalo2} here
we do not have an unloaded network representing free space, but
instead, a transition layer that enables electromagnetic waves
from free space to propagate into the TL network and vice versa.

\begin{figure} [h!]
\centering \epsfig{file=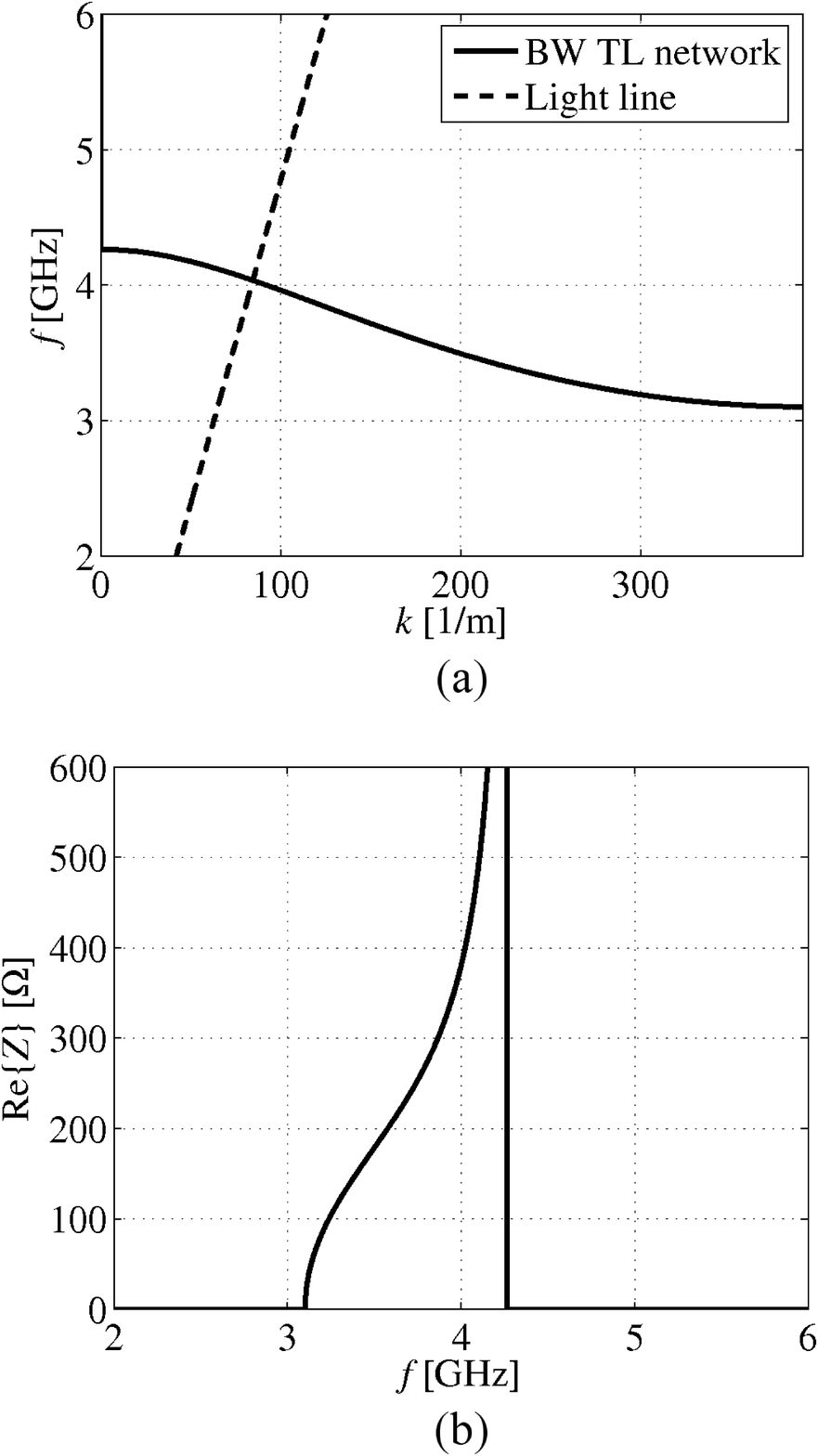, width=0.4\textwidth}
\caption{Dispersion (a) and impedance (b) in a three-dimensional
loaded transmission-line network with parameters as shown in
Table~I.} \label{dispersion_impedance}
\end{figure}

To optimize matching with free space, we need to design the TL
network in such a way that its impedance equals that of free space
(377~$\Omega$). Using the previously derived dispersion and
impedance equations,\cite{Alitalo1} we have found suitable
dimensions and parameters for the network operation in the
microwave region, see Table~I. The resulting dispersion and
impedance curves are presented in Fig.~\ref{dispersion_impedance}.
From Fig.~\ref{dispersion_impedance} we can conclude that the
optimal operation frequency for the network studied here is around
4~GHz. The operation frequency can be changed e.g. by varying the
values of the lumped capacitances and inductances.

\begin{table}[h!]
\centering \caption{Transmission-line network parameters.}
\label{table1}
\begin{tabular}{|c|c|c|c|c|}
\hline

Period & TL impedance & $C$ & $L$ & $\varepsilon_r$ \\

\hline

8~mm & $150$ $\Omega$ & 0.1~pF & 2.5~nH & 1\\

\hline
\end{tabular}
\end{table}

\section{Transition layer and simulation model}

\begin{figure} []
\centering \epsfig{file=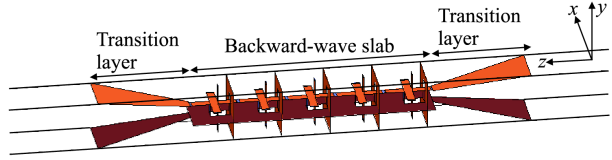, width=0.5\textwidth}
\caption{(Color Online). HFSS model of one ``unit cell'' of the
backward-wave slab.} \label{HFSS_model}
\end{figure}

As the TL network that is used here is similar to previous
designs\cite{Alitalo1,Alitalo2} it can be conveniently realized
using the microstrip technology. The transition layer can
therefore be realized with parallel-plate-waveguide type of TLs at
the ends of the network, as illustrated in Fig.~\ref{HFSS_model}.
Naturally, this way we can obtain operation for one polarization
only.\cite{Iyer} The benefit of this method is its simplicity. We
study a slab composed of the three-dimensional TL network
(infinite in the transversal directions in this case) and the TLs
of the transition layer cover the both surfaces of the slab.

We have made full-wave simulations of the proposed backward-wave
slab with the transition layers and sections of free space on both
sides of the slab. The simulations were done using Ansoft's High
Frequency Structure Simulator (HFSS). The simulation of the
transversally infinite slab can be greatly simplified by using
periodical boundary conditions. This way we can simulate only one
``unit cell'' of the slab, as shown in Fig.~\ref{HFSS_model}. The
TL network has the same parameters as shown in Table~I. The metal
strips are made of infinitely thin perfect conductors and have the
width of 1.3~mm with the distance from the ground being 2~mm.
Small holes are cut into the horizontal and vertical ground planes
to allow wave propagation in all axial directions. As shown in
Fig.~\ref{HFSS_model}, the thickness of the slab is five periods
in the direction of the $z$-axis. The width of the TLs of the
transition layer gradually changes from 2~mm to 7.988~mm, with
their length being 16~mm. The neighboring TLs of the transition
layer are not in contact, since there are approximately 12~$\mu
\rm m$ and 4~$\mu \rm m$ gaps between them in the $x$- and
$y$-directions, respectively.

\section{Simulation results}

\begin{figure} [b!]
\centering \epsfig{file=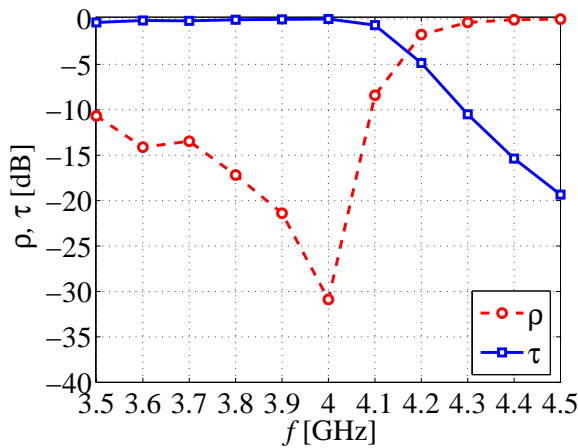, width=0.5\textwidth}
\caption{Simulated reflection and transmission through the
backward-wave slab as a function of frequency for the normal
incidence.} \label{rho_tau}
\end{figure}

First, the case of the normal incidence was studied, i.e., a plane
wave in free space with the wave vector $\overline{k}$ parallel to
the $z$-axis and electric field $\overline{E}$ parallel to the
$y$-axis illuminated the backward-wave slab. By studying the
reflection ($\rho$) and transmission ($\tau$) coefficients, it was
found that the optimal operation frequency for the structure was
approximately 3.6~GHz (the frequency point where most of the power
went through the slab). This implies that the impedance of the
network was best matched to free space at that frequency, since
the transition layer impedance does not depend on the frequency.
By changing the inductance value of the lumped inductances to
$L=2$~nH, the reflection and transmission curves shown in
Fig.~\ref{rho_tau} were obtained, showing good correspondence with
Fig.~\ref{dispersion_impedance}(b).

We believe that the slight difference in the inductance value with
which operation at 4~GHz is achieved, as compared to the
analytical study, is a result of differences between the
analytical and simulation models. The analytical equations do not
take into account the finite size of the lumped elements and the
equations used to calculate the data presented in
Figs.~\ref{dispersion_impedance}(a) and
\ref{dispersion_impedance}(b) do not take into account the finite
thickness of the slab.\cite{Alitalo1} The study of this inaccuracy
is out of the scope of this letter, since here we wish to
illustrate the feasibility of the proposed matching method as
such.

\begin{figure} []
\centering \epsfig{file=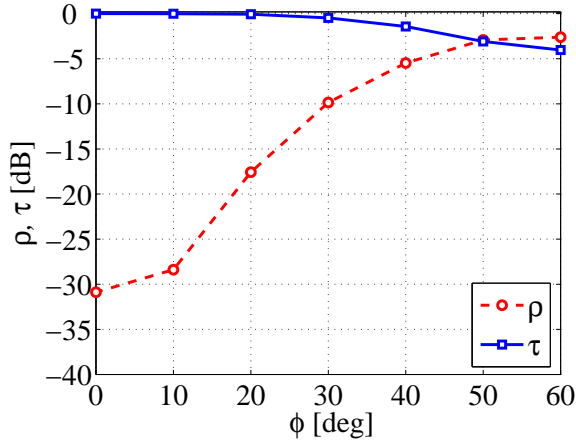, width=0.5\textwidth}
\caption{Simulated reflection and transmission through the
backward-wave slab as a function of the incidence angle at the
frequency 4~GHz.} \label{rho_tau_obl}
\end{figure}

From Fig.~\ref{rho_tau} it is seen that the optimal frequency of
operation (for normal incidence) is approximately 4~GHz. To study
the dependence of the transmission and reflection on the incidence
angle, we have made simulations with the same model for oblique
plane-wave illumination. The polarization of the illuminating wave
is kept the same, i.e., $\overline{E}$ is parallel to the $y$-axis
and therefore the incidence angle $\phi$ is the angle in the
$xz$-plane. See Fig.~\ref{rho_tau_obl} for the results when
$\phi=0^0...60^0$. We can conclude that the transition layer
operates very well also for fairly large oblique incidence angles,
although the transmission clearly reduces when $\phi$ grows.

\begin{figure} []
\centering \epsfig{file=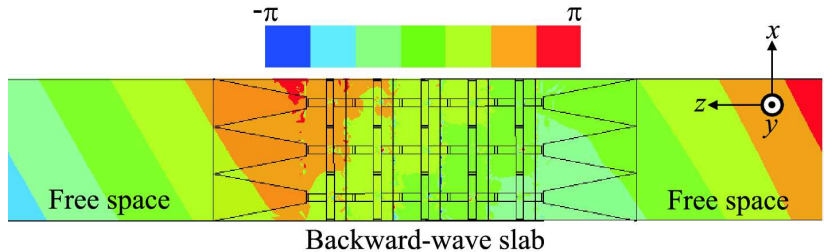, width=0.66\textwidth}
\caption{(Color Online). Simulated phase distribution in free
space and inside the backward-wave slab, at the frequency 4~GHz.
Incidence angle $\phi=30^0$.} \label{phase}
\end{figure}

With the model shown in Fig.~\ref{HFSS_model} the refraction on
the interface between free space and the backward-wave slab may
not be visible due to the small size of the model in the
transversal direction. To observe negative refraction, we have
extended the simulation model to encompass three ``unit cells'' of
the slab in the $x$-direction. See Fig.~\ref{phase} for the
simulated phase distribution in this larger simulation model at
the frequency 4~GHz and incidence angle of $\phi=30^0$. The
relative refractive index of the slab seems to be close to $-1$,
which is expected based on the dispersion curve in
Fig.~\ref{dispersion_impedance}(a).

Preliminary simulations show that at least for small angles of
incidence also in the $yz$-plane (angle $\theta$), the transition
layer operates well. For instance, for the incidence angle of
$\theta=20^0$ (with $\phi=0^0$) the reflection from the
backward-wave slab is below $-15$~dB at the frequency 4~GHz.

\section{Conclusions}

We have proposed and studied a transition layer for matching a
slab of a backward-wave transmission-line network with free space.
We have analytically studied how the network impedance can be
tuned to match that of free space. The proposed network and
transition layer have been simulated using a commercial full-wave
simulator. The simulation results show that the layer can be used
to match a backward-wave transmission-line network with free space
and verify the negative refraction on the interfaces between free
space and the network.

\section*{Acknowledgements}

This work has been partially funded by the Academy of Finland and
TEKES through the Center-of-Excellence program. Pekka Alitalo
wishes to thank the Graduate School in Electronics,
Telecommunications and Automation (GETA) and the Nokia Foundation
for financial support.

\end{document}